# Some features of effective radius and variance of dust particles in the numerical simulation of dust climate on Mars

**Chi-Fong Wong[1], Kim-Chiu Chow[1], Kwing L. Chan[1], Jing Xiao[1], and Yemeng Wang[1]**

[1]Space Science Institute / State Key Laboratory of Lunar and Planetary Science, Macau University of Science and Technology, Room A502, Avenida Wai Long, Taipa, Macau

Correspondence to: K. C. Chow, kcchow@must.edu.mo

**Key Points:**

- Analytical expressions of dust effective radius and variance have been derived for N-bin numerical schemes.
- Variations of effective radius and variance of dust particles have been investigated.
- The importance of variable effective radius in numerical simulation of Martian climate is discussed.

**Abstract**

Airborne dust is an important constituent in the Martian atmosphere because of its radiative interaction with the atmospheric circulation, and dust size is one crucial factor in determining this effect. In numerical modeling of the dust processes, description of the dust size is usually dependent on the choice of a particular size distribution function, or with fixed values of effective radius (ER) and effective variance (EV) though they are variable in reality. In this work, analytical expressions have been derived to specify ER and EV for N-bin dust schemes based on the model calculated dust mixing ratio. Numerical simulations based on this approach thus consider the effects of variable ER on the atmospheric radiation and their interaction. The results have revealed some interesting features of the dust distribution parameters such as the seasonal and spatial variation of ER and EV, which are generally consistent with some previous observational and modeling studies. Compared with the usual approach of using fixed ER, simulation results with the present approach suggest that the variability of ER can have significant effect on the simulated thermal field of the Martian atmosphere.

**Keywords**

Mars, dust, effective radius, effective variance, general circulation model

**1 Introduction**

Dust particles suspending in the atmosphere of Mars have an important effect on the Martian climate due to their radiative heating and cooling effects in the atmosphere (Gierasch and Goody, 1968, 1972). The radiative properties of dust are determined by the microscopic properties of the dust particles such as composition (i.e. complex index of reflection), shape, and size. While both composition and shape are usually assumed to be homogenous or uniform, the size distribution can be spatially and temporally varying, and so do the radiative effects. The effects of dust on radiation are basically dependent on the quantities Effective Radius (ER) and Effective Variance (EV). These two quantities generally represent the distribution of dust particles with different sizes in the atmosphere, and thus can be derived from the underlying size distribution if it is known (Hansen & Travis, 1974). Both values and variations of ER and EV are important to understanding the general circulation of Mars and so attracted the attention of some previous studies (e.g. Murphy et al., 1993; Kahre et al., 2008).

ER and EV have been measured by various remote-sensing observations since Mariner 9, Viking, and some subsequent missions (see reviews by Pollack et al. 1995; Tomasko et al. 1999; Dlugach et al. 2003; Smith 2008). These retrievals of ER require the assumption of some particular types of dust size distribution, such as the Gamma distribution (e.g. Lemmon et al. 2004; Wolff et al., 2006, 2009), the modified Gamma distribution (e.g. Clancy et al. 2003; Wolff and Clancy, 2003), and the Lognormal distribution (e.g. Fedorova et al. 2009, 2014). In these retrievals, EV is fixed and the value should be prescribed in priori in order to determine ER. In such a way consistent values of ER have been found to be around 1.5 μm (a canonical value), with EV chosen between 0.2 and 0.5. It is not surprising that seasonal or regional variation of ER has also been observed, first in Clancy et al. (2003) and Wolff and Clancy (2003), and recently in Smith et al.

(2016) and Vicente-Retortillo et al. (2017). It is found that ER is seemingly correlated to the atmospheric dust loading (see also Chen-Chen et al., 2019), with larger values of ER occur mostly in northern fall and winter. The vertical variations of ER have also been observed (Rannou et al., 2006; Fedorova et al., 2009; Clancy et al., 2010; Määttänen et al., 2013; Guzewich et al., 2014), indicating a strong gravitational segregation effect in most of the year except for the case when intense dust lifting events occur; during which large particles may reach a higher altitude and the effective dust size appears to be more uniform (see also Montmessin et al., 2017 and Kahre et al., 2017).

In many general circulation models (GCMs) of Mars, the dust cycle is usually simulated interactively in which dust lifting, transportation, and sedimentation processes are parameterized by some model-resolved parameters such as temperature, wind, and surface wind stress; while the radiative effects of the dust simulated may affect these parameters.

In some numerical models, the dust cycle is simulated by the two-moment scheme (e.g. Schulz et al., 1998; Morrison & Gettelman, 2008). By assuming a constant value of EV and a prescribed dust size distribution function, ER can be obtained from the model-calculated mass mixing ratio. The dust size distribution functions usually seen in literatures include the Gamma distribution (e.g. Lee et al., 2018) and the lognormal distribution (e.g. Madeleine et al., 2011; Wang et al., 2018).

In this work, we investigate the values and variations of ER and EV by using the Mars GCM MarsWRF with a two-particle scheme in which dust population is simulated by using two size bins. This kind of dust scheme is generally called a N-bin scheme, where dust particles are represented by a finite number of tracers of different particle sizes. Although the global dust cycle and its thermal effects have been reasonably simulated even with a small number of bins (e.g., Basu et al. 2004, 2006 for a two-bin case; Kahre et al. 2005, 2006 and Neary and Daerden 2018 for three-bin cases), so far there is no explicit formulation for the calculation of ER and EV in this kind of scheme, and fixed values of theses quantities should be given in priori. In this case, the interactive feature between the radiation and dust mixing ratio cannot be well represented when a fixed value of ER is used in radiation calculation. One objective of this study is to provide an analytical formulation for these two quantities for the usual N-bin schemes. Compared with the two-moment scheme, the present approach has some distinctive features. First, a priori knowledge about the dust size distribution is not needed, such as the functional form and value of EV. Second, the seasonal and spatial variations of EV can be obtained since EV is not assumed to be fixed. Third, the present approach can be readily extended to cases of arbitrary N.

The present study is focusing on the two-particle scheme since it is the simplest case when considering a varying ER. In fact, the analytical expressions introduced in the present study can be simply extended to the N-particle case. This simple scheme may indeed provide more insight on dust loading and its relationship with dust particle sizes. It is also worth mentioning that recent observations suggest the prevalence of a bimodal size distribution of dust (Montmessin et al., 2002, 2006; Määttänen et al., 2013; Fedorova et al., 2014), challenging the traditional assumption of a continuous monomodal distribution form. Therefore, the two-particle scheme focusing in this study is a suitable approach for a preliminary study on this topic.

In Section 2, the analytical expressions of ER and EV for the N-particle scheme will be derived. Particular attention will be focused on the case of N = 2 (two-particle scheme). In Section 3, numerical simulations performed by the general GCM MarsWRF will be described. The results of the simulations will be discussed in Section 4. Finally, the main results of the present study will be summarized and discussed in Section 5.

## 2 Mathematical formulations

2.1 General case (N-particle)

In a general N-particle scheme, dust is simulated by N particle types corresponding to radius $r_i$ with $i = 1, 2, \ldots, N$. The corresponding mass mixing ratios $q_{1,2,\ldots,N}$ of these types are traced individually and there are no interactions between different types of dust particles. At any grid point, the number density $n_i$ of particle type i is related to the mixing ratio $q_i = \rho_i/\rho_{CO_2} = (4/3)\pi r_i^3 n_i \rho_p/\rho_{CO_2}$, where $\rho_p$ and $\rho_{CO_2}$ are densities of the dust particle and the environmental $CO_2$ gas respectively. The particles are all assumed to be spherical in shape. Recall that for a continuous size distribution $n(r)$ with particle radius r, the kth-moment is given by the integral $M_k = \int_0^\infty r^k n(r) dr$. In the N-particle scheme, r takes discrete values and so the kth-moment can be given by a summation:

$$M_k = \sum_{i=1}^N r_i^k n_i = \sum_{i=1}^N r_i^{k-3} q_i \frac{3}{4\pi} \frac{\rho_{CO_2}}{\rho_p} = n_{tot} \frac{\sum_{i=1}^N q_i r_i^{k-3}}{\sum_{i=1}^N q_i r_i^{-3}}, \quad (1)$$

where the last equality is justified by the fact that $M_0 = \sum_{i=1}^N n_i = n_{tot} = \sum_{i=1}^N r_i^{-3} q_i (3/4\pi)(\rho_{CO_2}/\rho_p)$. As we can see, the kth-moment at any grid point is determined by the total number density $n_{tot}$ and the tracers $q_i$. Now, it is straightforward to express ER and EV for a N-particle scheme as they are both given by the moments of size distribution (Hansen & Travis, 1974):

$$r_{eff} \equiv \frac{M_3}{M_2} = \frac{\sum_{i=1}^N q_i}{\sum_{i=1}^N q_i r_i^{-1}}, \quad (2a)$$

$$v_{eff} \equiv \frac{M_4/M_3}{M_3/M_2} - 1 = \frac{(\sum_{i=1}^N q_i r_i)(\sum_{j=1}^N q_j r_j^{-1})}{(\sum_{i=1}^N q_i)^2} - 1, \quad (2b)$$

where $r_{eff}$ and $v_{eff}$ are ER and EV, respectively. Notice that the factor $n_{tot}$ is always cancelled out in the above expressions.

2.2 Two-particle case

In the two-particle scheme, using Eq. (2a), Eq. (2b) and taking N = 2 we have:

$$r_{eff} = \frac{q_1 + q_2}{q_1 r_1^{-1} + q_2 r_2^{-1}}, \quad (3a)$$

$$v_{eff} = \frac{q_1 q_2}{(q_1 + q_2)^2} \frac{(r_1 - r_2)^2}{r_1 r_2}. \quad (3b)$$

In these expressions both $r_{\text{eff}}$ and $v_{\text{eff}}$ can be fully determined by $q_1$ and $q_2$. Unlike the approach of continuous size distributions implemented in a two-moment scheme ($v_{\text{eff}}$ has to be fixed for solving $r_{\text{eff}}$ from tracers), both $r_{\text{eff}}$ and $v_{\text{eff}}$ in the above two-particle scheme can be solved independently and hence may provide a way to evaluate the spatial and temporal variation of $v_{\text{eff}}$.

The functional properties of $r_{\text{eff}}$ and $v_{\text{eff}}$ can be easily seen by noticing that both Eq. (3a) and Eq. (3b) can be converted to single variable functions. Let's define a dimensionless variable $R \equiv (q_1 - q_2)/(q_1 + q_2)$, that can be interpreted as the *relative abundance* among particle types at a particular grid point. For a fixed total mass mixing ratio of dust $q_{\text{tot}} = q_1 + q_2$, dust particles of the finer mode ($q_1$) may be more responsible for a large value of R while dust particles of the coarser mode ($q_2$) perform in the opposite way. In the extreme cases when $q_1 \gg q_2$ and $q_1 \ll q_2$, R may approach the limit of +1 and -1 respectively. In fact, $r_{\text{eff}}$ and $v_{\text{eff}}$ can also be converted to single variable functions by using the variable $q_2/q_1$ instead of R. However, the range of this variable will then become 0 to $\infty$. Now, in terms of $R$, we have

$$r_{\text{eff}} = \frac{2r_1 r_2}{[(r_1 + r_2) - (r_1 - r_2)R]}; \quad v_{\text{eff}} = (1 - R^2)\frac{(r_1 - r_2)^2}{4r_1 r_2}. \tag{4}$$

We can see that $r_{\text{eff}}$ is a monotonically decreasing function of R, while $v_{\text{eff}}$ behaves as a parabola centered at $R = 0$. The minimum and maximum values of $r_{\text{eff}}$ are respectively $r_1$ and $r_2$, corresponding to $R = +1$ and -1 respectively. This implies that $r_{\text{eff}}$ is close to the radius of dominant mode at the grid point when either $q_1$ or $q_2$ is negligible there. Also, when dust particles are generally larger, $r_{\text{eff}}$ will likely take a larger value. In this case, the minimum and maximum values of $v_{\text{eff}}$ are respectively 0 and $(r_1 - r_2)^2/(4r_1 r_2)$, corresponding to $R = \pm 1$ and 0 respectively.

### 3 Numerical model and simulations

To illustrate the quality of our approach, we utilize a robust numerical model MarsWRF to generate global circulation of the Martian atmosphere, especially its dust cycle. MarsWRF is the Mars-dedicated version of PlanetWRF (Richardson et al., 2007; Guo et al., 2009; Toigo et al., 2012), and was developed on the basis of the Weather Research and Forecasting (WRF) model for terrestrial weather and climate studies. MarsWRF is a grid point model utilizing Arakawa C-grid in horizontal directions, with 36 and 72 grids in latitudinal and longitudinal directions respectively, corresponding to the resolution of 5° squared or about 300 km squared in the equatorial region. The model has 52 terrain-following hydrostatic pressure layers, defined by surface pressure and a fixed model top pressure (taken as 0.0057 Pa). The basic model configuration and the physics schemes used are basically the same as those used in Chow et al. (2018) and Xiao et al. (2019).

As mentioned in the previous section, dust is simulated in this work by two size bins with particle radii of 1 and 3 μm in the numerical simulations. Processes of dust lifting, transportation, and sedimentation for each bin are controlled by model-resolved conditions. Dust lifting is parameterized by two schemes used in Newman and Richardson (2015) with some slight modifications (see below), and dust is assumed to be available everywhere and at all times over the whole planet surface except those surfaces with ice cover. In the first scheme the lifting of dust is function of surface wind stress. Dust lifting occurs over the surface when the local near-surface stress exceeds a particular threshold value (a constant value 0.043 N m-2). The second scheme parameterizes dust lifting due to thermal convection similar to dust devils. Dust lifting occurs when the temperature difference $\Delta T$ between surface and the surface air exceeds a certain threshold value (27K in this case). The amount of dust lifting is determined by thermodynamic efficiency $\Delta T/T_{\text{surf}}$, as well as the sensible heat flux. Dust lifted to the atmosphere is transported by the model-resolved wind, and then settling down under gravity according to the size-dependent Stokes–Cunningham relation.

The short-wave and long-wave radiation are evaluated by the Wide Band Model (WBM) as described in Richardson et al. (2007), which considers the radiative interaction of dust with the $CO_2$ atmosphere. Therefore, dust in the atmosphere may change the atmospheric radiation and thus the circulation. The contribution of dust to the long-wave and short-wave radiation is evaluated following Haberle et al. (1982) and Briegleb (1992) respectively. The change in the shortwave extinction opacity due to the suspending dust is given by (Madeleine et al., 2011):

$$\frac{d\tau}{dp} = \frac{3}{4g\rho_p} \frac{Q_{\text{ext}}}{r_{\text{eff}}} q_{\text{tot}}, \qquad (5)$$

where $g = 3.71\ m\ s^{-2}$ is the gravitational constant, $\rho_p$ is the density of dust particles, and $r_{\text{eff}}$ is the effective radius given by Eq. (3a). The extinction coefficient $Q_{\text{ext}}$ is considered as size-dependent. Its values are evaluated by the Python package *miepython*, which solves the Mie scattering theory numerically when a complex index of reflection is given. The values are then smoothed by a very narrow Gamma distribution of $v_{\text{eff}} = 0.02$ (Hansen & Travis, 1974, Figure 8). In this work, based on the observational data from Wolff and Clancy (2003) we adopt $Q_{\text{ext}} = 3.19$ and 2.92 for the particle sizes of 1 μm and 3 μm respectively. Besides, we keep other radiative parameters (e.g. single scattering albedo and asymmetric factor) size-independent for simplicity. This setting is reasonable as WBM only considers the averaged effects over a wide range of wavelength, and so is basically insensitive to the change of dust size. With a more sophisticated radiation scheme (e.g. correlated-k scheme), a full set of size-dependent parameters will be needed.

In this study, our primary goal is to explore the values and variations of ER and EV with Eqs. 3a and 3b. To evaluate the implementation of this approach in a Mars GCM, we perform a MarsWRF simulation (referred as *SimMain*) with the aforementioned setting for 16 Martian Years (MYs) and our main discussions will focus on the results from this simulation. Moreover, it is curious to ask whether the variability of ER is important in a Mars GCM simulation. For this purpose, we perform another 16 MY-simulation (referred as *SimRef*) as a reference, which has the same configuration as SimMain except using fixed values of $r_{\text{eff}} = 1.5$ μm and $Q_{\text{ext}} = 3.04$ for evaluating the dust opacity (Eq. 5). A comparison of SimMain and SimRef should reveal more information about the effect of variable ER in a Mars GCM simulation.

## 4 Results of simulations

In this section, the results from the two simulations SimMain and SimRef will be discussed. Both simulations are run for 16 MYs, with the first two MYs considered the spin-up time. As the main results of this work, dust seasons and dust size distribution of SimMain are discussed in Sec. 4.1 and 4.2. In Sec. 4.3, results from SimRef will be discussed and compared with SimMain.

4.1 Dust climate

In the 16 years of SimMain simulation, the dust climates in 13 years show a regular pattern (Fig. 1a) in which the simulated zonal-mean and column-integrated dust extinction optical depth (CDOD) is consistent with some derived observations (e.g., Madeleine et al., 2011; Montabone et al., 2015). For example, the dichotomy of the dust loading seasons can be captured; low-dust-loading (LDL) season in the northern spring and summer and high-dust-loading (HDL) season in the northern fall and winter. The simulation could also capture other prominent features of the dust climate including the two-episode feature in the equator region and the northern mid-latitude region (Xiao et al., 2019), and the corresponding solsticial pause around Ls = 270° (Lee et al., 2018). A *regular year* climate has been obtained by averaging the results of these 13 years.

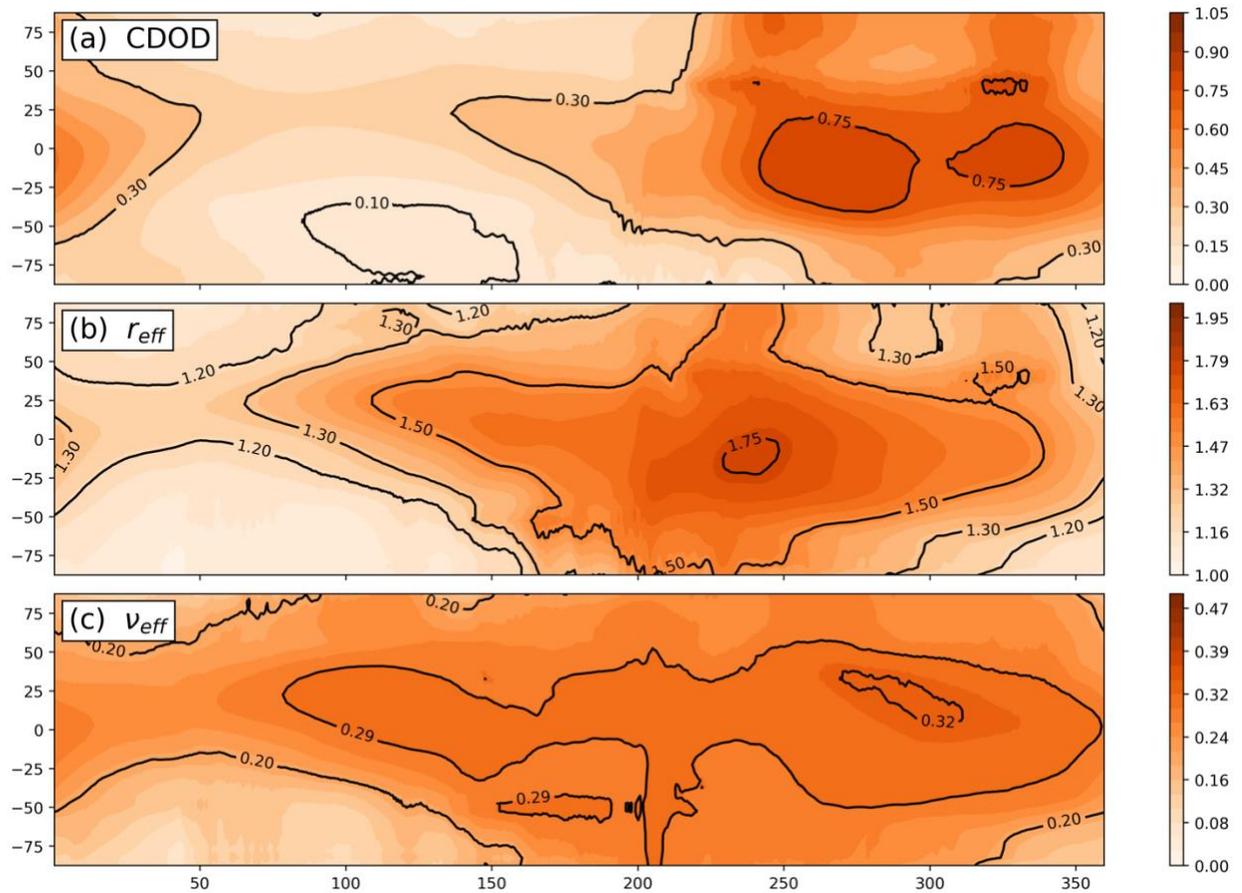

**Figure 1.** Annual variations of the zonal-mean column-integrated dust extinction optical depth in visible wavelength (0.67µm) (a), effective radius (b), and effective variance (c), given by the numerical simulation SimMain. Vertical and horizontal axes represent the latitude and solar longitude respectively.

## 4.2 Dust size

In Regular years, values and variations of zonal-mean and column-mean (defined by $\bar{x} = \int x \, dp/p_s$ for variable $x$, where $p_s$ is surface pressure and the integration is from the surface to the top of the atmosphere) ER and EV can be seen from Fig. 1b and 1c respectively. In the beginning of the year, ER can be as less as 1.1 µm (near polar regions), while it increases gradually to 1.3 µm around Ls = 60° (Fig. 1b). Its value reaches 1.5 µm around Ls = 100°, and reaches its annual peak of 1.75 µm around Ls = 240° in the equatorial region. This range of ER is consistent with the observed annual average of 1.5 µm. Also, we can see an approximate correlation between dust size and dust loading. There is another kind of dichotomy found in the seasonal variation of ER, in which large ER usually happens in HDL seasons. In contrast to the CDOD, ER presents only one peak at the equatorial region, while a two-peak pattern can still be found in the northern mid-latitudes.

Compared with ER, the variation of EV (Fig. 1c) seems relatively weaker. For most of the year, EV is found to be greater than 0.2 (except regions near the poles) and reaches a greater value about 0.29 from Ls = 90° to 360°. It reaches its peak around 0.32 from Ls = 260° to 310°. In general, EV is greater in this period. Unlike ER, there is neither an apparent dichotomy pattern for EV, nor a good correlation to the CDOD episodes. This feature is indeed consistent with the assumption of some observations, in which dust size is retrieved by prescribing a single-valued EV for the underlying size distribution. Also, a relatively uniform value of EV is qualitatively consistent with the previous model study (Kahre et al. 2008) where the EV seems universal.

Beside column-averaged quantities, vertical profiles may tell more about the spatial distribution of ER and EV. We consider two periods of Ls = 90° and Ls = 255°, which respectively represent the clearest and most dusty time in the Regular year climate. The zonal-mean vertical profiles of ER and EV at these times are shown in Fig. 2. Again, we observed that ER is generally smaller in the LDL season (Fig. 2a) and greater in the HDL season (Fig. 2b). In both periods, ER is larger near the ground and decreases as altitude increases. Large particle can extend to a greater height in low latitude regions compared with that in high latitude regions. These patterns show a strong gravitational segregation, and are qualitatively consistent with Madeleine et al. (2011) where dust size is simulated by two-moment scheme.

For EV, the general patterns in the vertical profiles are similar to those for ER. However, although values of EV in the LDL season (Fig. 2c) is generally smaller than those in the HDL season (Fig. 2d), the difference is not as apparent as that of ER. This result is indeed consistent with that in Fig. 1c. In both periods, greater values of EV can extend to a greater height in the low latitude region. At Ls = 90°, the maximum of EV occurs near the ground around the latitudes of 10° N to 50° N. However, at Ls = 255°, the maximum of EV (~ 0.32) occurs at the height of 25-30 km at the equatorial region. The altitudes of maximum EV decrease northward and southward from the equatorial region to form a bell shape (see Fig. 2d).

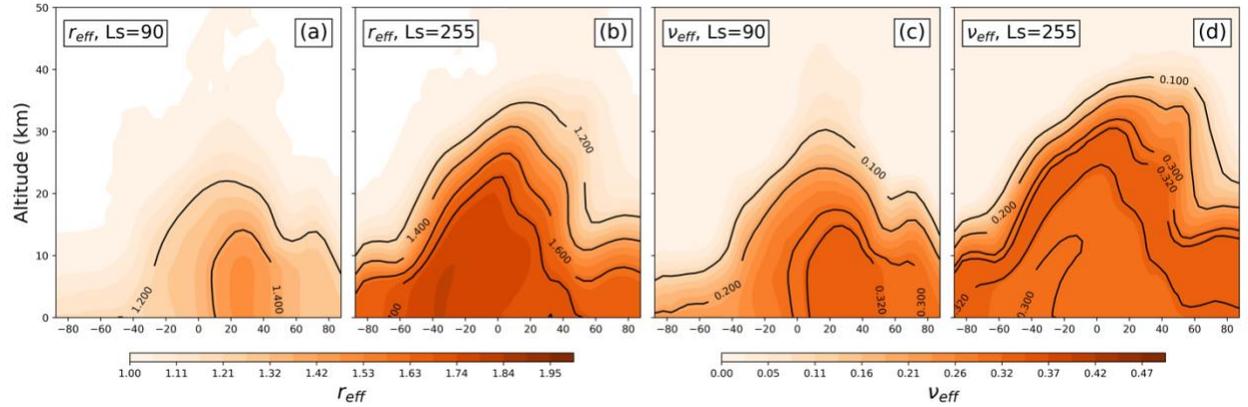

**Figure 2.** Zonal-mean vertical profiles of effective radius (a) (b) and effective variance (c) (d) at two periods of Ls = 90° (a) (c) and Ls = 255° (b) (d) in years with regular climate in SimMain. Horizontal and vertical axes represent the latitude and altitude respectively.

4.3 Effect on radiation process

It is interesting to ask what effect the variability of ER will bring about in a Mars GCM. The effect can be inspected by considering the difference in the simulated temperature field between SimMain and SimRef. The present discussion is based on the results from the regular year climate in SimMain and that in SimRef.

From the vertical profiles of the zonal-mean dust mixing ratio averaged over the latitudes from 40° S to 40° N (Fig. 3a), we can see the results are almost identical for SimMain and SimRef at Ls = 90°. On the other hand, for the corresponding values of ER (Fig. 3b) and EV (Fig. 3c), SimRef values are slightly larger than SimMain values. The situation is different at Ls = 255°. ER and EV are almost identical between the two simulations while the dust mixing ratios in SimMain (Fig. 3a) is apparently larger than that in SimRef.

Recall that dust opacity is proportional to factor $q_{tot}/r_{eff}$, where $r_{eff}$ is given by Eq. (3a) in SimMain, and is taken as 1.5 µm in SimRef. As height increases, the decreasing ER in SimMain pushes this factor greater thus generates the difference $\Delta(q_{tot}/r_{eff})$ between SimMain and SimRef. Moreover, the difference in dust mixing ratios between the two simulations would also contribute to $\Delta(q_{tot}/r_{eff})$. Fig. 3d shows the vertical profile of $\Delta(q_{tot}/r_{eff})$ at the two chosen time periods, calculated directly with the zonal- and latitudinal-mean values shown in Figs. 3a-c. From Fig. 3d we can see $\Delta(q_{tot}/r_{eff})$ reaches its maximum at the height about 20 km and 30 km at Ls = 90° and 255° respectively. Moreover, the maximal values at Ls = 255° is larger than that at Ls = 90°, mainly due to the difference in dust mixing ratio of the two simulations at that time.

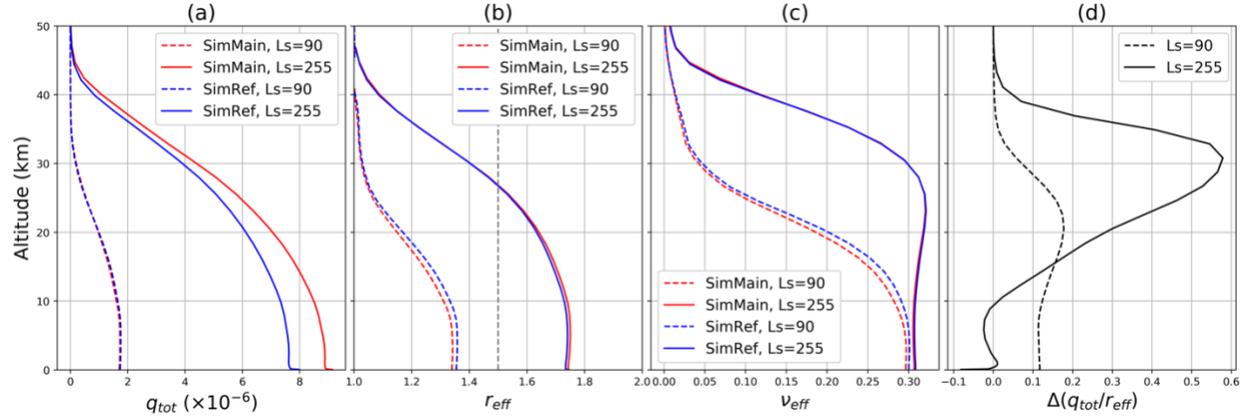

**Figure 3.** Vertical profiles of some zonal- and latitudinal-mean (40° S to 40° N) values at two time periods of Ls = 90° and 255° for the simulations SimMain and SimRef: (a) Total dust mixing ratio. (b) Effective radius. (c) Effective variance of, respectively. (d) Corresponding profiles for the difference in the factor $q_{tot}/r_{eff}$ between SimMain and SimRef. Although the GCM has resolved effective radius in SimRef (blue lines in Fig. 3b), calculations of opacity in SimRef use just 1.5 µm (gray dashed line in Fig. 3b), as mentioned in main text.

A difference in $q_{tot}/r_{eff}$ implies a difference in opacity, thus radiative heating rate, and thus the temperature field. The vertical profiles of the zonal-mean temperature differences between SimMain and SimRef at Ls = 90° and 255° are shown in Figs. 4a and 4b respectively. At Ls = 90°, there is an apparent band of positive temperature anomalies at the height from about 10 km to 30 km over the latitudinal region from 40° S to 40° N, with a maximum at about 20 km height. Similarly, at Ls = 255°, an even more apparent positive temperature anomalies occurs at the height from about 20 km to at least 60 km, with the maximum at the height between 35 km and 40 km for the same latitudinal region. These are warming effect generated by differences in $q_{tot}/r_{eff}$ indicated in Fig. 3d. The heights of maximums of $\Delta(q_{tot}/r_{eff})$ are consistent with the heights of maximums of temperature anomalies at both times. It demonstrates that whether including variation of dust size could affect the calculation on the thermal structure, revealing the use of fixed or variable ER could bring in significant alternation on the thermal structure in modeling.

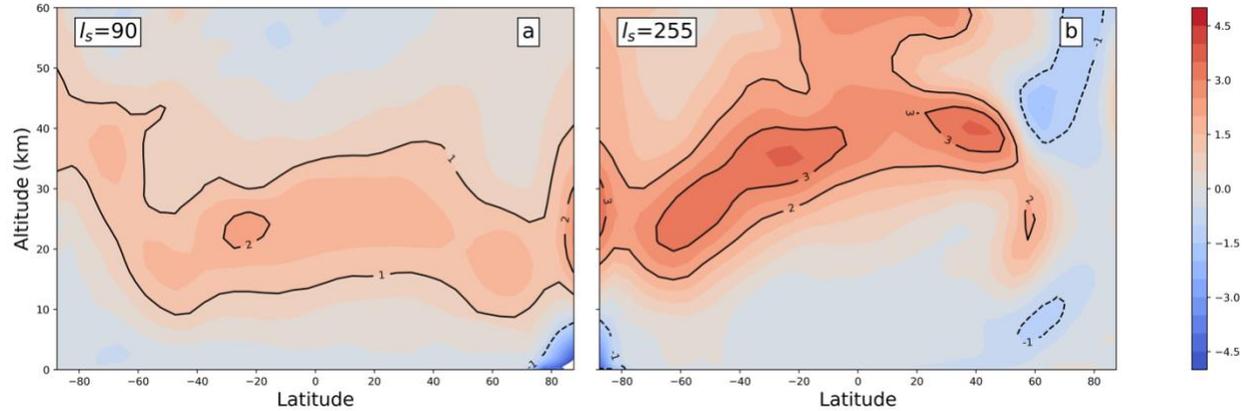

**Figure 4.** Vertical profiles of zonal-mean temperature difference (SimMain - SimRef) between the two simulations at (a) Ls = 90° and (b) Ls = 255°.

## 5 Discussions and Conclusions

Effective radius ER and effective variance EV are two important quantities concerning the distribution of dust particle sizes, and have been extensively studied as reported in many existing literatures. In most Mars GCMs, these quantities are commonly evaluated either by a two-moment scheme in which ER is related to the particle concentration through a prescribed distribution function, or by a N-bin dust scheme in which dust particles of N sizes are involved in the dust processes. The value of ER is generally chosen to be a fixed value in calculating the radiation associated with dust in N-bin schemes, although ER is a variable in reality. In this study, an approach to evaluate ER and EV has been proposed when a N-bin dust scheme is used in a Mars GCM. In this approach, ER and EV are given by analytical functions of the model-resolved dust mass mixing ratios. The functions are derived from the first principle and the information of the underlying size distribution and parameters is not required. In this approach, the effect of ER variation is taken into account during the calculation of radiation.

To illustrate the application of these formulations, we have performed numerical simulations with a GCM which considers dust particles of two sizes (two-particle scheme) and the dust particles are interactive with the radiation. Based on the present formulation, we obtain values and variations of ER which are consistent with previous observations and modeling studies.

Based on the simulation results with the present approach, the variation of EV has been evaluated. The values of EV are usually assumed to be constant in many GCMs. We find that the seasonal variation of EV has the general pattern similar to that of ER, but is relatively less apparent. In most of the year, EV is not significantly changed at some observed locations on Mars. This uniformity of EV is usually assumed in most literatures, and this property has been further supported by the present study.

To investigate the significance of variable ER in numerical simulations, we have compared two simulations. One adopts a variable ER (from the present formulation) for opacity calculation while another one adopts a fixed valued of ER. The results suggest that due to the vertical variation of ER (and increase of dust loading), an additional heating is present at the height between 10 and 30 km in the LDL season of Ls = 90° (height between

20 and 50 km in the HDL season at Ls = 255°) in the case with a variable ER. This effect is similar to that reported in the literature.

The present study is a preliminary investigation on the problem of dust size distribution and its interactive effect on the dust radiative process. Some parameters in the radiative process such as the scattering albedo, asymmetric factor and extinction coefficient could be dust-size dependent but have been considered as size independent in this study. A more sophisticated investigation in the future should consider the size-dependent nature of these parameters. Furthermore, a larger number of size bins should also be considered in future studies.

**Acknowledgments**

This research is funded by the grants from the FDCT of Macau (grant nos. 080/2015/A3 and 0088/2018/A3). MarsWRF was obtained from the PlanetWRF modeling project (Richardson et al., 2007; Toigo et al., 2012), which is a collaboration between Aeolis Research, Cornell, JPL, and the University of Toronto, and funded by numerous NASA grants. The data of the simulations can be accessed at https://figshare.com/account/home#/collections/4428632.

**References**

Basu, S., Richardson, M., & Wilson, R. (2004). Simulation of the Martian dust cycle with the GFDL Mars GCM. Journal of Geophysical Research E: Planets, 109(11), 1-25. doi: 10.1029/2004JE002243

Basu, S., Wilson, J., Richardson, M., & Ingersoll, A. (2006). Simulation of spontaneous and variable global dust storms with the GFDL Mars GCM. Journal of Geophysical Research E: Planets, 111(9). doi: 10.1029/2005JE002660

Briegleb, B. (1992). Delta-Eddington approximation for solar radiation in the NCAR community climate model. Journal of Geophysical Research, 97(D7), 7603-7612. doi: 10.1029/92JD00291

Chen-Chen, H., Perez-Hoyos, S., & Sanchez-Lavega, A. (2019, February). Dust particle size and optical depth on Mars retrieved by the MSL navigation cameras. Icarus, 319, 43-57. doi: 10.1016/j.icarus.2018.09.010

Chow, K.-C., Chan, K.-L., & Xiao, J. (2018). Dust activity over the Hellas basin of Mars during the period of southern spring equinox. Icarus, 311, 306-316. doi: 10.1016/j.icarus.2018.04.011

Dlugach, Z., Korablev, O., Morozhenko, A., Moroz, V., Petrova, E., & Rodin, A. (2003). Physical Properties of Dust in the Martian Atmosphere: Analysis of Contradictions and Possible Ways of Their Resolution. Solar System Research, 37(1), 1-19. doi: 10.1023/A:1022395404115

Fedorova, A., Montmessin, F., Rodin, A., Korablev, O., Määttänen, A., Maltagliati, L., & Bertaux, J.-L. (2014). Evidence for a bimodal size distribution for the suspended aerosol particles on mars. Icarus, 231, 239-260. doi: 10.1016/j.icarus.2013.12.015

Gierasch, P. J., & Goody, R. M. (1968). A study of the thermal and dynamical structure of the martian lower atmosphere. Planetary and Space Science, 16(5), 615-646. doi: 10.1016/0032-0633(68)90102-5

Gierasch, P. J., & Goody, R. M. (1972). The Effect of Dust on the Temperature of the Martian Atmosphere. Journal of the Atmospheric Sciences, 29(2), 400-402.

Guo, X., Lawson, W. G., Richardson, M. I., & Toigo, A. (2009, July). Fit-ting the Viking lander surface pressure cycle with a Mars General Circulation Model. Journal of Geophysical Research-Planets, 114, E07006. doi: 10.1029/2008JE003302

Haberle, R., Leovy, C., & Pollack, J. (1982). Some effects of global dust storms on the atmospheric circulation of Mars. Icarus, 50(2-3), 322-367. doi: 10.1016/ 0019-1035(82)90129-4

Hansen, J., & Travis, L. (1974). Light scattering in planetary atmospheres. Space Science Reviews, 16(4), 527-610. doi: 10.1007/BF00168069

Kahre, M., Hollingsworth, J., Haberle, R., & Murphy, J. (2008). Investigations of the variability of dust particle sizes in the martian atmosphere using the NASA Ames General Circulation Model. Icarus, 195(2), 576-597. doi: 10.1016/j.icarus.2008.01.023

Kahre, M., Murphy, J., & Haberle, R. (2006). Modelling the Martian dust cycle and surface dust reservoirs with the NASA Ames general circulation model. Journal of Geophysical Research E: Planets, 111(6). doi: 10.1029/2005JE002588

Kahre, M., Murphy, J., Haberle, R., Montmessin, F., & Schaeffer, J. (2005). Simulating the Martian dust cycle with a finite surface dust reservoir. Geophysical Research Letters, 32(20), 1-5. doi: 10.1029/2005GL023495

Kahre, M., Murphy, J., Newman, C., Wilson, R., Cantor, B., Lemmon, M., & Wolff, M. (2017). The Mars dust cycle. In the Atmosphere and Climate of Mars (p. 295-337). doi: 10.1017/9781139060172.010

Lee, C., Richardson, M. I., Newman, C. E., & Mischna, M. A. (2018, September). The sensitivity of solsticial pauses to atmospheric ice and dust in the MarsWRF General Circulation Model. Icarus, 311, 23-34. doi: 10.1016/j.icarus.2018.03.019

Määttänen, A., Listowski, C., Montmessin, F., Maltagliati, L., Reberac, A., Joly, L., & Bertaux, J.-L. (2013). A complete climatology of the aerosol vertical distribution on Mars from MEx/SPICAM UV solar occultations. Icarus, 223(2), 892-941. doi: 10.1016/j.icarus.2012.12.001

Madeleine, J.-B., Forget, F., Millour, E., Montabone, L., & Wolff, M. (2011). Re-visiting the radiative impact of dust on Mars using the LMD Global Climate Model. Journal of Geophysical Research E: Planets, 116(11). doi: 10.1029/2011JE003855

Montabone, L., Forget, F., Millour, E., Wilson, R., Lewis, S., Cantor, B., Wolff, M. (2015). Eight-year climatology of dust optical depth on Mars. Icarus, 251, 65-95. doi: 10.1016/j.icarus.2014.12.034

Montmessin, F., Korablev, O., Lefèvre, F., Bertaux, J.-L., Fedorova, A., Trokhimovskiy, A., Chapron, N. (2017). SPICAM on Mars Express: A 10 year in-depth survey of the Martian atmosphere. Icarus, 297, 195-216. doi: 10.1016/j.icarus.2017.06.022

Montmessin, F., Quémerais, E., Bertaux, J., Korablev, O., Rannou, P., & Lebonnois, S. (2006). Stellar occultations at UV wavelengths by the SPICAM instrument: Retrieval and analysis of Martian haze profiles. Journal of Geophysical Research E: Planets, 111(9). doi: 10.1029/2005JE002662

Montmessin, F., Rannou, P., & Cabane, M. (2002, June). New insights into Martian dust distribution and water-ice cloud microphysics. Journal of Geophysical Research-Planets, 107(E6), 5037. doi: 10.1029/2001JE001520

Morrison, H., & Gettelman, A. (2008). A new two-moment bulk stratiform cloud microphysics scheme in the community atmosphere model, version 3 (CAM3). Part I: Description and numerical tests. Journal of Climate, 21(15), 3642-3659. doi: 10.1175/2008JCLI2105.1

Neary, L., & Daerden, F. (2018). The GEM-Mars general circulation model for Mars: Description and evaluation. Icarus, 300, 458-476. doi: 10.1016/j.icarus .2017.09.028

Newman, C., & Richardson, M. (2015). The impact of surface dust source exhaustion on the martian dust cycle, dust storms and interannual variability, as simulated by the MarsWRF General Circulation Model. Icarus, 257, 47-87. doi: 10.1016/j.icarus.2015.03.030

Pollack, J., Ockert-Bell, M., & Shepard, M. (1995). Viking Lander image analysis of Martian atmospheric dust. Journal of Geophysical Research, 100(E3), 5235-5250. doi: 10.1029/94JE02640

Richardson, M., Toigo, A., & Newman, C. (2007). PlanetWRF: A general purpose, local to global numerical model for planetary atmospheric and climate dynamics. Journal of Geophysical Research E: Planets, 112(9). doi: 10.1029/2006JE002825

Schulz, M., Balkanski, Y. J., Guelle, W., & Dulac, F. (1998, May). Role of aerosol size distribution and source location in a three-dimensional simulation of a Saharan dust episode tested against satellite-derived optical thickness. Journal of Geophysical Research-Atmospheres, 103(D9), 10579-10592. doi: 10.1029/97JD02779

Smith, M. D. (2008). Spacecraft observations of the Martian atmosphere. Annual Review of Earth and Planetary Sciences, 36, 191-219. (WOS:000256391900008) doi: 10.1146/annurev.earth.36.031207.124335

Toigo, A., Lee, C., Newman, C., & Richardson, M. (2012). The impact of resolution on the dynamics of the martian global atmosphere: Varying resolution studies with the MarsWRF GCM. Icarus, 221(1), 276-288. doi: 10.1016/j.icarus.2012.07.020

Tomasko, M., Doose, L., Lemmon, M., Smith, P., & Wegryn, E. (1999). Proper-ties of dust in the Martian atmosphere from the Imager on Mars Pathfinder. Journal of Geophysical Research E: Planets, 104(E4), 8987-9007. doi: 10.1029/1998JE900016

Vicente-Retortillo, A., Martínez, G., Rennó, N., Lemmon, M., & de la Torre-Juárez, M. (2017). Determination of dust aerosol particle size at Gale Crater using REMS UVS and Mastcam measurements. Geophysical Research Letters, 44(8), 3502-3508. doi: 10.1002/2017GL072589

Wang, C., Forget, F., Bertrand, T., Spiga, A., Millour, E., & Navarro, T. (2018). Parameterization of Rocket Dust Storms on Mars in the LMD Martian GCM: Modeling Details and Validation. Journal of Geophysical Research: Planets, 123(4), 982-1000. doi: 10.1002/2017JE005255

Wolff, M. J., & Clancy, R. T. (2003, September). Constraints on the size of Martian aerosols from Thermal Emission Spectrometer observations. Journal of Geo-physical Research-Planets, 108(E9), 5097. doi: 10 .1029/2003JE002057

Xiao, J., Chow, K.-C., & Chan, K.-L. (2019). Dynamical processes of dust lifting in the northern mid-latitude region of Mars during the dust storm season. Icarus, 317, 94-103. doi: 10.1016/j.icarus.2018.07.020